\documentstyle[12pt]{article}
\begin{document}
\begin{center}
\hfill  IITB-TPH-0001\\
\vspace{1.6cm}
\centerline{\bf Baryogenesis through axion domain wall}
\vspace{0.5cm}
\centerline{S.N. Nayak and U.A.Yajnik}
\vspace{0.2cm}
\centerline{Physics Department, Indian Institute of Technology Bombay, Mumbai-  400076, India}
\end{center}
\bigskip
\vspace{0.2cm}
\noindent
\begin{center}
{\bf Abstract}
\end{center}
 
Generic axion models give rise to axion domain walls in the early
Universe and they have to disappear not to overclose the
universe, thus limiting the nature of discrete symmetry allowed
in these type of models. 
Through QCD sphalerons, net chiral charge can be created by these
collapsing walls which in turn can result the observed baryon assymetry.

\vspace{9cm}
nayaks@phy.iitb.ernet.in

yajnik@phy.iitb.ernet.in
\vfill

\newpage
The explanation of observed baryon asymmetry of the universe is still
a hunting problem in the realm of cosmology and particle physics,
thus generating lots of diverse ideas and activity. The earlier scenario
to produce the asymmetry via the decay of heavy gauge bosons 
and scalar \cite{re1} in
GUTS cannot survive the sphaleron wash out at the electroweak scale,
unless B-L is an exact symmetry \cite{re2}. The scenario of electroweak baryogenesis
through sphaleron transition also runs into problem  because of inadequete
CP violation in Higgs sector and more importantly, it is  not clear
that the electroweak phase transition is strongly  first order to 
realise  the out of equilibrium condition through bubble dynamics.
In this background it was suggested to generate baryon asymmetry
through topological defects (the remnants of some earlier  symmetry
breaking) at the electroweak scale \cite{re3}. In this scenario 
the baryogenesis takes place inside the core of the defects where
the sphaleron transition takes place.
Here we discuss the
issue in the context of axion domain wall and show that we can
produce sufficient amount of baryons at the scale much below
the weak scale. Similar situation has been considered recently
by Brandenberger  et al \cite{re4}.

Many axion models \cite{re5} also have discrete Z(N) symmetry which is
spontaneously broken at $T = \Lambda_{QCD}$. 
This is generic for any axion models where the Pecci-Quinn symmetry
$U_{PQ}(1)$ is broken only by QCD gluon anomaly. In the above N is the
number of quark flavours that rotate under $U_{PQ}(1)$.
Because of this 
discrete symmetry, there  exist N degenerate and distinct $CP$
conserving minima of the  axion potential   which is of the  form
\begin{equation}
V(a) = {m_a}^2 ({v_{PQ}} /N)^2  [1-f(aN / {v_{PQ}})],
\end{equation}
where $f$ is a periodic function of period $2\pi$ and $v_{PQ}$ is the
Pecci-Quinn scale.
These disconnected  and degenerate  vacuum states gives rise to
axion  domain walls  at $T = \Lambda_{QCD}$,  when  the  discrete
symmetry  is spontaneously broken.
The  resulting  domain walls have thickness  $\Delta = {m^{-1}_a}$
and  surface energy  density $\eta =  m_a {v^{2}}_{PQ}$, where $m_a$ is the
mass of axion.

These domain walls are  disastrous  cosmologically \cite{zeld} and has to
disappear  so that they do  not  over close the universe, unless N=1. One way
to achieve  this  is to  introduce a soft breaking  term  
of  the  form  $ \mu^3 \Phi$ \cite{re6}. 
Here we are considering DFSZ axion model and $\Phi$ is the singlet
under standard gauge group \cite{dfs}.
This would produce an effective
value of $\theta_{QCD}$ of order $\mu^3 / {{m_a}^2 v_{PQ}}$.
For this to be consistent with the upper limit on the electric dipole
moment of neutron we get
\begin{equation}
\mu^3 \leq 10^{-9} \frac {{f_\pi}^2 {m_\pi}^2}{v_{PQ}}.
\end{equation}

This soft breaking term would produce a shift in energy  density among the 
degenerate vacuum, hence a pressure towards the  domain with
highest vacuum energy leading to annihilation of walls.
It is also possible that the domain walls created may not
survive the QCD phase transition since $Z(N)$ symmetry may 
be dynamically broken.

In this letter we argue that even for the brief period that they
exist they can produce sufficient amount of baryons. 
The important ingredient that goes in to our argument is the existence
of the sphaleron like configuration in QCD and the rate of this topological
transition is given by \cite{re7}
\begin{equation}
\Gamma_S = \kappa {\alpha_s}^4 T^4.
\end{equation}
In the above $\alpha_s$ is the strong coupling constant and the 
proportionality constant $\kappa$ can be of the order thousand \cite{pr}. This
is the transition rate over the  potential energy barrier
separating vacua of different Chern-Simons number. But unlike
the  electoweak case  where the sphaleron transition is the source
of baryon number violation; the QCD sphaleron does not  induce  any
baryon number violation  since  it  has  only parity conserving vector
couplings. So as it is, this scenario will create some chiral  charge
separation  mechanism. Baryogenesis will be achieved if additionally we 
have a nonvanishing chemical potential induced by other mechanism. For
instance it could be a background field effect as in spontaneous baryogenesis
scenario and its variants \cite{re8,tur}.

The CP violating phase that is needed for baryogensis is nothing
but the strong CP violating parameter $\theta_{QCD}$ which need
not be zero at high temperature. The value of $\theta$ has to be
decided by  some  stocastic process in a given horizon  volume. The
model we are discussing where the domain wall has to disapear
due to the explicit soft  breaking term has an effective $\theta$
that is consistent with above experimental
constraint and also ensures
that the walls do not overclose the universe \cite{zeld}.
We take the  CP violating phase to be of order $10^{-10}$.

The final ingredient for the baryogensis is the departure from
thermal equlibrium. In our scenario this is automatically
achieved when the walls annihilate due to difference in
vacuum energy. 
The situation is similar to the model independent pictures of defect
mediated baryogenesis. Whereas mere translational motion or long
lived defects cannot induce any net assymmetry, collapse and mutual
annihilation can lead to creation of net assymmetry \cite{re3}.
Let $V_{BG}$ be the effective three dimensional volume in which the
time irreversible processes occur during the disappearance of walls.
Then the net baryon number density is then given by
\begin{equation}
\Delta n_B = \frac{1}{V} \frac{\Gamma_S}{T} V_{BG} \Delta \theta,
\end{equation}
with V as the total
volume.

As discussed earlier, the above formula needs to be supplemented
by the contribution from a mechanism that converts the net chiral
charge into baryonic charge. For example, 
one can consider an extra factor ${m_f}/T$ with
$m_f$ as the fermion mass \cite{tur}, in evaluating the baryon 
number density. In the baryogenesis scenario where electroweak sphaleron
transitions takes place in the core of topological defects,
this factor turns out to be order one. In our case this factor can
enhance the rate of baryon production since the temperature
we are interested is of QCD scale. But at present we are not considering  
this factor, as we are presenting our picture in a qualitative way and
one has to see whether it enters in our calculation or not.
Then the baryon to entropy ratio in volume V is
\begin{equation}
\frac{\Delta n_B}{s} = {g^*}^{-1} {\alpha_s}^4 {\Delta\theta} \frac{V_{BG}}{V}.
\end{equation}
To evaluate the volume suppression factor, let us take the
average  separation  of the domain walls as $\xi(t)$, which 
from kibble mechanism is
\begin{equation}
\xi(t) = {T_c}^{-1},
\end{equation}
where $T_c$ is the  temperature where  Z(N) symmetry  is
spontaneously  broken  and is equal to $ (m_a v_{PQ})^{1/2}$.
Then the volume
occupied  by the domain walls in a horizon size $d_H (t)$
is
\begin{equation}
V_{BG}  = {\xi (t)}^ 2  {{m_a}^{-1}} (\frac{d_{H(t)}}{\xi (t)})^3.
\end{equation}  
The last factor is the number of domains in the horizon
volume.  With this the volume suppression factor turns out to be
\begin{equation}
\frac{V_{BG}}{V} =  { (m_\pi f_\pi)^{1/2} / m_a}. 
\end{equation}
In the  above we have  used $T_c = ({m_a v_{PQ}})^{1/2} 
= ({f_\pi m_\pi})^{1/2}$.
Since at QCD scale the pion mass can go to zero the above
volume suppression factor can be of order unity for suitable value
of the axion mass. This aspect of the problem requires detailed calculation
in the specific axion model. But qualitatively, the  
thickness of the axion wall goes
inversly proportional to the axion mass. The mass of the axion
due to instanton effect at
QCD scale is 
\cite{re9}
\begin{equation}
m_a (T) = 0.1 m_a (T= 0) (\Lambda_{QCD} / T )^{3.7}.
\end{equation}
So it is possible that at temperature just around QCD scale the
thickness of the wall is only a few order of magnitude smaller than
the horizon size and ${V_{BG}}/V$ need not be a serious suppression factor.

Another crucial
criteria that our picture satisfies, is the fitting of QCD sphaleron 
inside the axion domain wall, hence requiring 
no modification in the bulk value of $\Gamma_S$.
The size of QCD sphaleron will be of order
${\Lambda^{-1}}_{QCD}$ which is smaller than 
the wall thickness that is ${m_a}^{-1}$
for allowed value of axion mass.
So with $\kappa$ of order thousand, the CP  violating phase
of the order  $10^{-10}$  and  $g_*$ is of the order $10$ we can
produce sufficient amount of  baryons  at the QCD  scale.

\vspace{1cm}
\begin{center}
{\bf Acknowledgements}
\end{center}

This work was carried out as a part of a Department of Science and Technology
project SP/S2/K-08/97.


\vspace{1cm}

\begin{thebibliography}{99}
\frenchspacing

\bibitem{re1} S. Dimopoulos and L. Suskind, Phys. Rev. D18 (1978) 4500;
M. Yoshimura, Phys. Rev. Lett 41 (1978) 281; A. Ignatiev, N. Krasnikov,
V. Kuzumin and A. Tavkheelidze, Phys. Lett. B76 (1978) 436; 
S. Weinberg, Phys. Rev. Lett. 42 (1977) 850;
D. Toussaint, S. Trieman, F. Wilczek and A. Zee, Phys. Rev. D19 (1979) 1036.

\bibitem{re2} V. Kuzmin, V. Rubakov and M. Shaposhnikov,
Phys. Lett. B155 (1985) 36. 

\bibitem{re3} R. Brandenberger, A.C. Davis, T. Prokopec and M. Trodden,
Phys.Rev. D53 (1996) 4257; 
R. Brandenberger, A.C. Davis, and M. Trodden,
Phys. Lett. B335 (1994) 123; 
S.Duari and A.U.Yajnik,
Phys. Lett. B326 (1994) 212. 

\bibitem{re4} R. Brandenberger, I. Halperin and A. Zhitnitsky,
hep-ph/9808471 and 9903318, to be published in the proceedings of SEWM-98.

\bibitem{re5} M.S. Turner, Phys. Rept. 197 (1990) 67; G.G. Raffelt, Phys. Rept.
198 (1990) 1. 

\bibitem{zeld} Y.B. Zel'dovich, I.Y. Kobzarev and L.B. Okun, Zh. Eksp. Teor. Fiz
67 (1974) 3.

\bibitem{re6} P.Sikivie,
Phys. Rev. Lett. 48 (1982) 1150. 

\bibitem{dfs} M. Dine, W. Fischler and M. Srednicki, Phys. Lett. 
B104 (1981) 199;
A.P. Zhitnitskii, Sov. J. Nucl. Phys. 31 (1980) 260.

\bibitem{re7} L. Mclerran, E. Mottola and M.E. Shaposhnikov,
Phys. Rev. D43 (1990) 2027.

\bibitem{pr} P. Arnold and L. McLerran, Phys. 
Rev.D36 (1987) 581, D37 (1988) 1020.

\bibitem{re8} S. Dodelson and L.M. Widrow,
Phys. Rev. D42 (1990) 326; A. Cohen, D. Kaplan and A. Nelson, Phys. Lett.
B263 (1991) 86 and Nucl. Phys B373 (1992) 453.

\bibitem{tur} N. Turok and T. Zadrozny, Phys. Rev. Lett 65 (1990) 2331,
Nucl. Phys. B358 (1991) 451; L. Mclerran, M. Shaposhnikov, N. Turok and
M. Voloshin, Phys. Lett. B256 (1991) 451.

\bibitem{re9} D. Gross, R. Pisarski and L. Yaffe, Rev. Mod. Phys. 53 (1981) 43;
J. Preskill, M.B. Wise and F. Wilczek, Phys. Lett. B120 (1983) 127;
L.F. Abbot and P. Sikivie, Phys. Lett. B120 (1983) 133;
M. Dine and W. Fischler, Phys. Lett. B120 (1983) 137.

\end{thebibliography}
\end{document}